%% file: Roche.tex
\documentclass[useAMS,usenatbib,intlimits,centertags]{mn2e}
\usepackage{graphicx}
\usepackage{amsmath}
\usepackage{booktabs}
\usepackage{times}

\allowdisplaybreaks

\renewcommand{\d}{\text{d}}

\begin{document}

 \title{A Roche Model for Uniformly Rotating Rings}

 \author[S.\ Horatschek \& D.\ Petroff]
 {Stefan Horatschek\footnotemark\addtocounter{footnote}{-1}
 and David Petroff\thanks{E-mail: {\tt S.Horatschek@tpi.uni-jena.de} (SH);\newline
 {\tt D.Petroff@tpi.uni-jena.de} (DP)}\\
 Theoretisch-Physikalisches Institut, University of Jena, Max-Wien-Platz 1, 07743 Jena, Germany}
 \date{\today}

 \pagerange{\pageref{firstpage}--\pageref{lastpage}} \pubyear{2009}

 \maketitle

 \label{firstpage}

 \begin{abstract}
  \input{Roche_abstract}
 \end{abstract}


\input{Roche_content}

 \bibliographystyle{mn2e}
 \bibliography{Reflink}

\appendix

\input{Roche_appendix}

 \label{lastpage}

\end{document}

%% file: Roche_abstract.tex
A Roche model for describing uniformly rotating rings is presented and the results are compared with numerical
solutions to the full problem for polytropic rings.
In the thin ring limit, the surfaces of constant pressure including the surface of the ring itself
are given in analytic terms, even in the mass-shedding case.

%% file: Roche_content.tex
\section{Introduction}
A uniformly rotating star with a sufficiently soft equation of state can
be described approximately using the Roche model \citep{Roche73}. In this model the matter
is treated as a test fluid in a $1/r$-potential, i.e.\ one considers the
whole mass of the star to be concentrated at the centre. By doing so,
self-gravitating effects of the outer mass shells are neglected.
Such Roche models have been considered, especially in the mass-shedding limit,
by \citet{ZN71,SS02}. For polytropes with indices $n\ga 2.5$ this approximation yields
results that differ by less than about a percent from their correct values \citep{Meineletal08}.
Whereas most analytical solutions to figures of equilibrium describe bodies with
constant mass density (e.g.\ the Maclaurin spheroids, Jacobi ellipsoids, etc.),
the Roche model is very useful because it is applicable to non-homogeneous matter.
\par

Aside from spheroidal figures of equilibrium, there exist configurations with toroidal topology.
Such rings have been studied both analytically \citep{Kowalewsky85,Poincare85,Poincare85b,Poincare85c,
Dyson92,Dyson93,Ostriker64b,PH08b} and numerically \citep{Wong74,ES81,EH85,Hachisu86,AKM03,FHA05}.
\par

Here we apply the basic idea of the Roche model to rings. Thus, we do not choose to have
the test fluid rotate in the field of a point mass, but in that
of a circular line of mass with constant linear mass density. In addition
to the mass, we thus also have to specify the radius of the circle.
\par

For the comparison of the solutions of the Roche model with those of the full problem for
polytropes, we use a multi-domain spectral program, much like the one described in \citet{AKM03b}
and a similar one tailored to Newtonian bodies with toroidal topologies (see \citealt{AP05} for more information).

\section{The Roche Model with a Ring Source}
\subsection{Basic Equations}

If we use cylindrical coordinates $(\varrho,z,\varphi)$, then the gravitational potential
of a circular line centred on the axis of mass $M$ and radius $b$ reads
\begin{align}
 U &= -\frac{GM}{\pi}\int_0^\pi\!\frac{\d\varphi}{\sqrt{b^2+\varrho^2+z^2-2b\varrho\cos\varphi}}\\
   &= -\frac{2GM}{\pi\sqrt {(b+\varrho)^2+z^2}}
        \,K\!\left(\frac{2\sqrt{b\varrho}}{\sqrt{(b+\varrho)^2+z^2}}\right),\label{pot}
\end{align}
where $G$ is the gravitational constant and $K$ denotes the complete elliptic integral of the first kind,
\begin{align}
 K(k):=\int_0^{\frac{\pi}2}\frac{\d\theta }{\sqrt{1-k^2\sin^2\theta}}.
\end{align}
Sometimes it will be convenient to use the polar-like coordinates $r$ and
$\chi$ defined by
\begin{align}\label{r_chi}
 \varrho = b - r\cos\chi \quad\text{and}\quad z = r\sin\chi.
\end{align}
For a test fluid rotating uniformly with the angular velocity $\Omega$, Euler's equation
can be written as
\begin{align}\label{Euler}
 \nabla\left(U+h-\frac12\Omega^2\varrho^2\right)=0,
\end{align}
where $h$ is defined by
\begin{align}\label{h}
 h:=\int_0^p\frac{\d p'}{\mu(p')},
\end{align}
and where $p$ is the pressure, $\mu$ the mass density and $U$ the potential given above%
\footnote{For isentropic matter, $h$ is simply the specific enthalpy.}.
We integrate \eqref{Euler} and get
\begin{align}\label{int_Euler}
  U+h-\frac{1}{2}\Omega^2\varrho^2=V_0,
\end{align}
where $V_0$ is the constant of integration. Evaluating this equation
at the surface of the ring, along which the function $h$ vanishes, then leads to
\begin{align}\label{int_Euler_surf}
  U_\text{s}-\left.\frac{1}{2}\Omega^2\varrho^2\right|_\text{s}=V_0.
\end{align}
Via this equation, the surface of the ring,
described by $r_\text s=r_\text s(\chi)$ or $z_\text s=z_\text s(\varrho)$, and
thus the ratio of inner and outer radius
$\varrho_\text i$ and $\varrho_\text o$
\begin{align}
 A:=\frac{\varrho_\text i}{\varrho_\text o},
\end{align}
are given implicitly for prescribed $M$, $b$, $\Omega^2$ and $\varrho_\text o$. Analogously,
equation \eqref{int_Euler} can be used to find surfaces of constant $h$.
\par

In what follows, we simplify the equations by introducing the dimensionless quantities
\begin{subequations}\label{dimensionless}
 \begin{align}
  \frac{\bar r}{r}=\frac{\bar b}{b}=\frac{\bar\varrho}{\varrho}=\frac{\bar z}{z}=\frac{1}{\varrho_\text o},\\
  \frac{\bar\Omega^2}{\Omega^2}=\frac{\varrho_\text o^3}{GM},\quad
  \frac{\bar U}{U}=\frac{\bar V_0}{V_0}=\frac{\bar h}{h}=\frac{\varrho_\text o}{GM},
 \end{align}
\end{subequations}
which implies $\bar\varrho_\text i=A$ and $\bar\varrho_\text o=1$.
In the dimensionless versions of the above equations, $GM$ cancels out.
Except for two scaling constants (e.g.\ $M$ and $\varrho_\text o$), two parameters
are necessary to describe a ring in the Roche model (e.g.\ $\bar b$ and $\bar\Omega^2$).

\subsection{Mass-Shedding Configurations}

Of particular interest are configurations at the mass-shedding limit. In this limit,
a fluid particle at the outer rim rotates with the Kepler frequency, which means that
the gravitational force is balanced by the centrifugal force alone -- the pressure gradient vanishes.
For the squared angular velocity at the mass-shedding limit we find
\begin{align}\label{ms}
\bar\Omega_\text{ms}^2=\left.\frac{\partial\bar U}{\partial\bar\varrho}\right|_{\bar\varrho=1,\bar z=0}.
\end{align}
With the potential \eqref{pot} this gives
\begin{align}
\bar\Omega_\text{ms}^2= \frac{2E\!\left(\bar b\right)}{\pi\left(1-\bar b^2\right)},
\end{align}
where $E$ denotes the complete elliptic integral of the second kind,
\begin{align}
E(k):=\int_0^{\frac{\pi}2}\sqrt{1-k^2\sin^2\theta}\,\d\theta.
\end{align}
Contrary to the general case, only one parameter is free, the other one is fixed by equation
\eqref{ms}.
\par

Mass-shedding configurations have a cusp at the outer rim%
\footnote{One can easily show that a configuration with a cusp must be at the mass-shedding limit.
The converse statement, that mass-shedding configurations have a cusp, can be proved if one makes the very reasonable
assumption that the pressure does not increase for increasing values of $z$ in the upper half-space \citep{Paehtz07}.},
and we will now calculate the associated angle.
At the surface, the quantity
\begin{align}
 V := U-\frac12\Omega^2\varrho^2
\end{align}
is constant, see \eqref{int_Euler_surf}, which means that along the surface we have
\begin{align}
 0=\frac{\d^2V}{\d\varrho^2} = V_{,\varrho\varrho} + 2V_{,\varrho z}\frac{\d z_\text s}{\d\varrho}
 +V_{,zz}\left(\frac{\d z_\text s}{\d\varrho}\right)^{\!2} + V_{,z}\frac{\d^2 z_\text s}{\d\varrho^2},
\end{align}
where a comma indicates partial differentiation.
The equatorial symmetry implies that first derivatives with respect to $z$
vanish at $z=0$, thus leading to
\begin{align}
\varrho=\varrho_\text o,\ z=0: \quad \frac{\d z_\text s(\varrho)}{\d\varrho}=\sqrt{-\frac{V_{,\varrho\varrho}}{V_{,zz}}}
=\sqrt{\frac{\Omega_\text{ms}^2- U_{,\varrho\varrho}}{U_{,zz}}}.
\end{align}
Hence the full angle is
\begin{align}\label{delta}
\begin{split}
\delta&=\left.2\arctan\sqrt{\frac{\Omega_\text{ms}^2- U_{,\varrho\varrho}}{U_{,zz}}}\right|_{\varrho=\varrho_\text o,\ z=0}\\
      &=1-\frac{2\left(1-\bar b^2\right)E\!\left(\bar b\right)}{(1-\bar b^2)K\!\left(\bar b\right)-2E\!\left(\bar b\right)},
\end{split}
\end{align}
cf.\ Fig.~\ref{delta_bild}.

\subsection{The Shape of the Surface}
To treat both mass-shedding configurations and the general case, we parametrize the angular velocity by
\begin{align}\label{Om2}
 \Omega^2 = \alpha\Omega^2_\text{ms},\qquad \alpha\in[0,1].
\end{align}
This choice of $\alpha$ arises from the fact that the mass-shedding limit
($\alpha=1$) poses an upper bound for the angular velocity and a solution
to \eqref{int_Euler_surf} can be found for arbitrarily small $\Omega^2$.
\par

Evaluating \eqref{int_Euler_surf} at the point $(\bar\varrho=\bar\varrho_\text o=1,\bar z=0)$ yields
\begin{align} \label{V0o}
 \bar V_0 = -\frac{1}{\pi}\left[\frac{\alpha E\!\left(\bar b\right)}{1-\bar b^2}+2K\!\left(\bar b\right)\right]
\end{align}
and at the point $(\bar\varrho=\bar\varrho_\text i=A,\bar z=0)$ gives
\begin{align} \label{V0i}
 \bar V_0 = -\frac{1}{\pi}\left[\frac{\alpha A^2E\!\left(\bar b\right)}{1-\bar b^2}+\frac{2}{\bar b}K\!\left(\frac{A}{\bar b}\right)\right].
\end{align}
Requiring that both expressions for $V_0$ agree leads to
\begin{align}\label{V0o_V0i}
 K\!\left(\bar b\right) = \frac{1}{\bar b}K\!\left(\frac{A}{\bar b}\right)
     - \frac{\alpha\left(1-A^2\right)E\!\left(\bar b\right)}{2\left(1-\bar b^2\right)}.
\end{align}
At an arbitrary surface point $(\bar\varrho,\bar z_\text s(\bar\varrho))$ we get
\begin{align}\label{V0arbitrary}
\begin{aligned}
  \bar V_0 = -\frac{1}{\pi}\Bigg[ &
   \frac{\alpha \bar\varrho^2E\!\left(\bar b\right)}{1-\bar b^2}+  \\ &
   \frac{2}{\pi\sqrt{\left(\bar b+\bar\varrho\right)^2+\bar z_\text s^2}}
      \,K\!\left(\frac{2\sqrt{\bar b\bar\varrho}}{\sqrt{\left(\bar b+\bar\varrho\right)^2+\bar z_\text s^2}}\right)\Bigg].
\end{aligned}
\end{align}
Together with \eqref{V0o} this is an implicit equation for the surface function $\bar z_\text{s}(\bar\varrho)$
or $\bar r_\text s(\chi)$. Unfortunately this function cannot be found in analytic terms,
however it is not difficult to handle it numerically. Furthermore, it will be treated analytically
for thin rings in the next subsection.
\par

For a prescribed $\alpha$, the requirement $A\ge 0$ means that \eqref{V0o_V0i} can only be satisfied for $\bar b$
larger than some minimal value.
It turns out that smaller values of $\bar b$ no longer describe rings, but spheroidal figures.
In this case, \eqref{V0i} and \eqref{V0o_V0i} must be replaced by
\begin{align}
 \bar\varrho=0,\ \bar z=\bar z_\text p: \qquad \bar V_0 = -\frac{1}{\sqrt{\bar b^2 + \bar z_\text p^2}}
\end{align}
and
\begin{align}
 K\!\left(\bar b\right) = \frac{\pi}{2\sqrt{\bar b^2 + \bar z_\text p^2}}
     - \frac{\alpha E\!\left(\bar b\right)}{2\left(1-\bar b^2\right)},
\end{align}
where $\bar z_\text p=z_\text p/\varrho_\text o$ denotes the dimensionless polar radius, in other words, the
ratio of the polar to the equatorial radius. The transition from spheroidal to toroidal topologies is described
when $\bar z_\text p=0$ or equivalently $A=0$. In the limit $\bar b\to 0$, the potential \eqref{pot} becomes
that of a point mass $M$ at the star's centre and one recovers the `standard' Roche model, cf.\ Appendix~\ref{standard_Roche}.

\subsection{Thin Rings}
\begin{figure*}
\centerline{\includegraphics{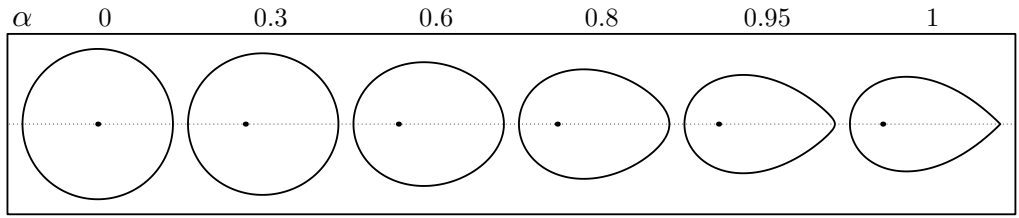}}
\caption{Cross-sections of rings in the thin ring limit scaled such that the horizontal extent of each is equal.
         The function $s_1(\chi)$ is depicted for various values of $\alpha$ in polar coordinates, where the
         axis of rotation is to the left of each cross-section and infinitely far away.
         The dots indicate the origin, where the source of the potential is located.\label{surf_s}}
\end{figure*}

In the thin ring limit in which $A\to 1$, clearly $\bar b$ also tends to 1.
This limit is of particular interest, especially since
a self-gravitating thin ring is also tractable to analytical methods \citep{Kowalewsky85,Poincare85b,Poincare85c,
Poincare85d,Dyson92,Dyson93,Ostriker64b,PH08b}.
By expanding the surface function
\begin{align}\label{rs_sum}
 \bar r_\text s(\chi) = \sum_{i=1}^\infty s_i(\chi)\left(1-\bar b\right)^i
\end{align}
about the thin ring limit $\bar b\to 1$, equation \eqref{int_Euler_surf} yields
\begin{align}
 0 = \alpha[1+s_1(\chi)\cos\chi] + \ln s_1(\chi),
\end{align}
which can be solved by using the Lambert W function, which fulfils the equation $W(x)\text e^{W(x)}=x$:
\begin{align}
 s_1(\chi) = \text e^{-W(\alpha\text e^{-\alpha}\cos\chi)-\alpha}
           = \frac{W(\alpha\text e^{-\alpha}\cos\chi)}{\alpha\cos\chi}.
\end{align}
For $\alpha\to 0$, we find $s_1(\chi)=1$ meaning that the cross-section tends toward a circle for $\bar b\to1$.
For $\alpha=1$ (and only for this value), $s_1(\chi)$ is
not differentiable at the point $\chi=\pi$ although both one-sided derivatives exist. One finds
\begin{align}
 \lim_{\chi\uparrow\pi}\frac{\d s_1(\chi)}{\d\chi}=-\lim_{\chi\downarrow\pi}\frac{\d s_1(\chi)}{\d\chi}=1,
\end{align}
which implies that the angle at this point is $\delta=\pi/2$, which can also be derived from \eqref{delta}
in the limit $\bar b\to1$. A series of pictures showing the shapes of
$s_1(\chi)$ for various values of $\alpha$ can be found in Fig.~\ref{surf_s}.
Higher terms $s_i(\chi)$ from the series \eqref{rs_sum} can be found iteratively and contain
powers of $\ln\left(1-\bar b\right)$.
\par

\begin{figure}
\centerline{\includegraphics{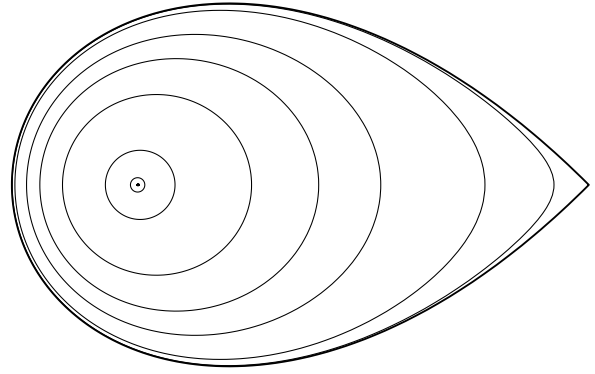}}
\caption{Lines of constant pressure (corresponding to constant $\bar h$) are shown for the mass-shedding ring
 in the thin ring limit. The function $t_1^{\bar h}(\chi)$ with $\alpha=1$ is depicted for various values
 of $\bar h$ in polar coordinates, where the axis of rotation is to the left of the cross-section and infinitely far away.
 The surface, which corresponds to $\bar h=0$, is precisely $s_1(\chi)$
 and can also be found in Fig.~\ref{surf_s}. The values of $\bar h$ starting
 from the surface and moving inward are 0, 0.001, 0.01, 0.05, 0.1, 0.2, 0.5 and 1. The value of
 $\bar h$ tends to infinity at the point, where the source is located.
 \label{consth}}
\end{figure}

To calculate surfaces of constant $h$, $\bar r_{\bar h}(\chi)$,
we can easily generalize the above calculation. Like \eqref{rs_sum}, we expand
these surfaces about the thin ring limit
\begin{align}
\bar r_{\bar h}(\chi) = \sum_{i=1}^\infty t^{\bar h}_i(\chi)\left(1-\bar b\right)^i.
\end{align}
As mentioned above, the function $h$ vanishes at the surface, i.e.\ $\bar r_\text{s}(\chi)=\bar r_0(\chi)$
and $s_i=t^0_i$. Equation \eqref{Euler} then gives
\begin{align}
 0 = \alpha\left[1+t_1^{\bar h}(\chi)\cos\chi\right] + \ln t_1^{\bar h}(\chi) + \pi\bar h,
\end{align}
and thus
\begin{align}
 t_1^{\bar h}(\chi) = \text e^{-W(\alpha\text e^{-\alpha-\pi\bar h}\cos\chi)-\alpha-\pi\bar h}
                    = \frac{W(\alpha\text e^{-\alpha-\pi\bar h}\cos\chi)}{\alpha\cos\chi}.
\end{align}
This function is depicted in Fig.~\ref{consth} for $\alpha=1$ and various values of $\bar h$.
For large values of $\bar h$, we have
\begin{align}
 \lim_{\bar h\to\infty} t_1^{\bar h}(\chi)\text e^{\alpha+\pi\bar h} =1
\end{align}
meaning that the curves of constant $h$ become circular as the source is approached.
In the limit $\alpha\to 0$ we find that
\begin{align}\label{ht}
h=-\frac{GM}{b}\frac1{\pi}\ln t,\qquad t\in(0,1]
\end{align}
only depends on the coordinate $t:=\bar r/(1-\bar b)$, but not on $\chi$.

\section{Comparison with the Full Problem}
The comparison of the Roche model with the full problem requires the identification of
various physical quantities. Whereas $M$, $\Omega$, $V_0$, $A$ and $z_\text s(\varrho)$
are clearly defined in both,
the value of $b$, which describes the location of the singular source in the Roche model, is not defined in
the full problem. There, we choose $b$ to be the position of the centre of mass of a meridional
cross-section of the ring.
\par

The Roche model contains two scaling
parameters, which can be `eliminated' by using the dimensionless quantities introduced
in \eqref{dimensionless}, and two physical parameters. On the other hand, for a given
equation of state, the full problem is determined by specifying a scaling parameter and
only one physical parameter. When comparing a specific Roche model with a solution to the full problem, one thus has
freedom as to how to make such a comparison. One can, for example, choose to have $\bar b$
and $\bar\Omega$ agree and then compare $\bar V_0$, $A$ and the shape of the surface.
One could also choose to have $\bar b$ and $A$ agree and then compare $\bar\Omega$, $\bar V_0$
and the shape of the surface.
\par

There must be some well-defined way of choosing the additional parameter in the Roche model if
the full problem is to tend to it in some limit. Let us consider rings made up of matter
obeying the equation of state
\begin{align}
 p=K\mu^{1+1/n},
\end{align}
i.e.\ polytropes, where $K$ is the polytropic constant and $n$ the polytropic index.
If such rings are expanded about the limit $A\to 1$, then a solution results,
the first few terms of which provide a good approximation to the full problem over some range
of values for $A$ \citep{Ostriker64b, PH08b}. This range of values shrinks to the single point $A=1$
in the limit $n\to\infty$, i.e.\ in the `isothermal limit'.
This means that the `isothermal' rings must be infinitely thin. Because we expect that only such rings
can be treated arbitrarily well in the Roche model, we do have a well defined way of choosing the
`additional parameter': namely $A=1$.
The one remaining free physical parameter, e.g.\ $\alpha$, corresponds to the
single physical parameter one has in the full solution and interpolates between mass-shed rings and
those with circular cross-sections. We thus expect that there exist isothermal (and necessarily
infinitely thin) rings with non-circular cross-sections, cf.\ Fig.~\ref{surf_s}.
\par

Those with circular cross-sections were studied in the papers mentioned in the last paragraph
and the results presented here must coincide with those for $\alpha=0$.
Indeed expression (148) for the potential and (35) with (151) for the angular velocity
in \citet{Ostriker64b} together with $\beta^{-1}\gg\xi\gg 1$
show that $\Omega^2 b^2$ is negligible compared to $U$, as implied by $\alpha=0$ and \eqref{Om2}.
Furthermore, the logarithmic behaviour \eqref{ht} can be recovered from (120), (122), (125)
and (127) in \citet{Ostriker64b}%
\footnote{Equation (122) in \citet{Ostriker64b} should read $\rho=\lambda\text e^{-\Psi}=\rho_0\text e^{-\Psi}$.}
\citep[see also][]{Ostriker64, PH08b}
\begin{align}\label{hOstriker}
h_\text{iso}=h_\text{c}-2K\ln\left(1+\frac{\xi^2}8\right),\qquad \xi\in[0,\infty).
\end{align}
In expression \eqref{ht} the function $h$ diverges at the centre $t=0$ and the surface is located
at the finite radius $t=1$, whereas in \eqref{hOstriker} the corresponding function remains finite at
the centre $\xi=0$, but diverges at the surface $\xi\to\infty$. To compare the expressions for $h$,
we thus take a derivative in order to eliminate the physically irrelevant constant
and consider only the regular portion of the $t$-interval, $t>0$, which corresponds to infinite values
of $\xi$. This leads to
\begin{align}
 t\frac{\d h}{\d t} = -\frac{GM}{\pi b}
\end{align}
and
\begin{align}
 \lim_{\xi\to\infty}\xi\frac{\d h_\text{iso}}{\d\xi} = -4K,
\end{align}
which can indeed be seen to be in agreement upon taking into account
\begin{align}
 GM = 4\pi K b,
\end{align}
see (146) and the definitions (97) in \cite{PH08b}. The situation is analogous to that in the spherically symmetric case
as discussed in Appendix~\ref{standard_Roche}.
\par

If we depart from the thin ring limit, then
the comparison with numerical solutions to the full problem of a uniformly rotating,
self-gravitating ring allows us to address the particularly important question of
how good the Roche model is. Tables~\ref{convergence_n_A70} and \ref{convergence_n_ms} show
how this model converges toward the full solution for polytropic rings as the
polytropic index $n$ is increased. The first of these provides a comparison of
rings with the radius ratio $A=0.7$ and the second for mass-shedding
configurations ($\alpha=1$ for the Roche model). In both cases, the value of $\bar b$ for the Roche model was chosen to agree
with the numerical one. The second column, $\kappa$, gives
an indication of how concentrated the mass is. If we consider the cross-section of a ring, and
define the two points in the equatorial plane at which the
density falls off to half of its maximal value
\begin{align}
 \mu(\varrho_\text a,z=0)=\mu(\varrho_\text b,z=0)=\frac{\mu_\text{max}}{2},
  \quad \varrho_\text b>\varrho_\text a,
\end{align}
then $\kappa$ is defined to be the ratio of
the distance between these two points to the total width of the ring's cross-section
\begin{align}\label{kappa}
 \kappa:=\frac{\varrho_\text b -\varrho_\text a}{\varrho_\text o -\varrho_\text i}.
\end{align}
\begin{table}\centering
\caption{Comparison between polytropic rings with $A=0.7$ and the corresponding Roche configuration with the same $\bar b$.
$\Delta$ denotes the difference between the values of the Roche configuration and the values of the full solution.
\label{convergence_n_A70}}
\begin{tabular}{ccccccc}\toprule
 $n$  &     $\kappa$        &       $\bar b$       &   $\Delta\Omega/\Omega_\text{num}$   &    $\Delta V_0/V_{0\,\text{num}}$ \\ \midrule
$1.0$ & $6.3\times 10^{-1}$ & $8.48\times 10^{-1}$ & $2.6\times 10^{-2}$ & $7.4\times 10^{-3}$ \\
$1.5$ & $4.9\times 10^{-1}$ & $8.46\times 10^{-1}$ & $1.9\times 10^{-2}$ & $6.0\times 10^{-3}$ \\
$3.0$ & $2.7\times 10^{-1}$ & $8.42\times 10^{-1}$ & $8.2\times 10^{-3}$ & $3.1\times 10^{-3}$ \\
$5.0$ & $1.3\times 10^{-1}$ & $8.34\times 10^{-1}$ & $2.6\times 10^{-3}$ & $1.2\times 10^{-3}$ \\
$7.0$ & $7.2\times 10^{-2}$ & $8.26\times 10^{-1}$ & $8.6\times 10^{-4}$ & $4.7\times 10^{-4}$ \\ \bottomrule
\end{tabular}
\end{table}%
\begin{table*}\centering
\caption{Comparison between polytropic rings in mass-shedding and the corresponding Roche configuration with the same $\bar b$.
$\Delta$ denotes the difference between the values of the Roche configuration and the values of the full solution.
\label{convergence_n_ms}}
\begin{tabular}{ccccccc}\toprule
 $n$  &     $\kappa$        &       $\bar b$       &     $\Delta A/A_\text{num}$    &   $\Delta\Omega/\Omega_\text{num}$   &   $\Delta\delta/\delta_\text{num}$   &   $\Delta V_0/V_{0\,\text{num}}$ \\ \midrule
$1.0$ & $5.3\times 10^{-1}$ & $5.77\times 10^{-1}$ & $-3.7\times 10^{-1}$ & $-3.8\times 10^{-2}$ & $2.7\times 10^{-2}$ & $-4.1\times 10^{-2}$ \\
$1.5$ & $4.0\times 10^{-1}$ & $6.04\times 10^{-1}$ & $-1.0\times 10^{-1}$ & $-2.0\times 10^{-2}$ & $1.3\times 10^{-2}$ & $-2.1\times 10^{-2}$ \\
$3.0$ & $2.1\times 10^{-1}$ & $6.53\times 10^{-1}$ & $-1.2\times 10^{-2}$ & $-3.9\times 10^{-3}$ & $2.1\times 10^{-3}$ & $-4.4\times 10^{-3}$ \\
$5.0$ & $1.1\times 10^{-1}$ & $6.92\times 10^{-1}$ & $-2.0\times 10^{-3}$ & $-7.7\times 10^{-4}$ & $3.4\times 10^{-4}$ & $-8.8\times 10^{-4}$ \\
$7.0$ & $5.7\times 10^{-2}$ & $7.20\times 10^{-1}$ & $-4.7\times 10^{-4}$ & $-2.0\times 10^{-4}$ & $7.8\times 10^{-5}$ & $-2.4\times 10^{-4}$ \\ \bottomrule
\end{tabular}
\end{table*}%
Only for $\kappa\ll 1$ is the Roche model expected to give good results. We find that $\kappa\la 0.2$ for polytropic rings with
$n\ga 3$. Such polytropes always have values for the radius ratio $A\ga0.45$, which implies that
Roche models with a smaller radius ratio do not provide a particularly close approximation to
any toroidal polytrope.
\par

\begin{figure}
\centerline{\includegraphics{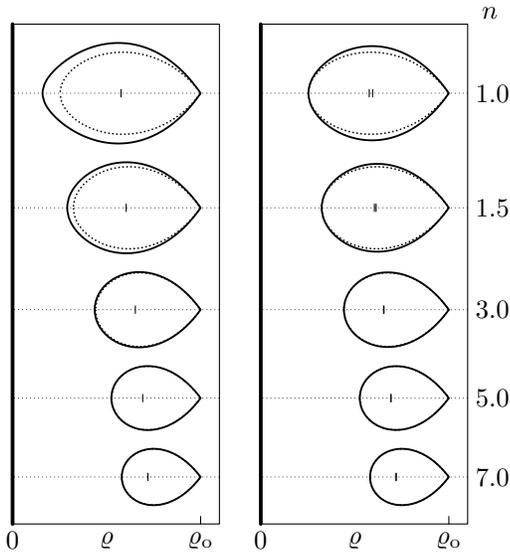}}
\caption{Comparison between the cross-sections of polytropic rings at the mass-shedding limit (dotted lines)
 and those of the corresponding Roche configurations, i.e.\ with $\alpha=1$ (solid lines). In the left panels
 the Roche model was chosen such that $\bar b$ agrees with the numerical value, whereas in the
 right panels, the values of $A$ were chosen to coincide. The ticks in the `centre' indicate the
 values of $\bar b$. When $A$ is prescribed (right panels), it turns out that $\bar b_\text{num}
 <\bar b_\text{Roche}$.
 \label{surf_ms}}
\end{figure}

The shapes of cross-sections of mass-shedding rings for various polytropic indices are
plotted in Fig.~\ref{surf_ms}. Here, the differences that arise due to different prescriptions
of the parameters are evident for small values of the polytropic index. When $A$ is chosen to
coincide between the Roche model and the full solution, then the two surfaces do not differ
appreciably, even for $n=1$. The radius ratio itself, and consequently the shape, is quite
different for a Roche model with $\bar b$ as prescribed from the solution to the full problem
with $n=1$. These differences vanish as $n$ is increased and the two prescriptive choices converge
to the full solution as they must.
\par

\begin{figure}
\centerline{\includegraphics{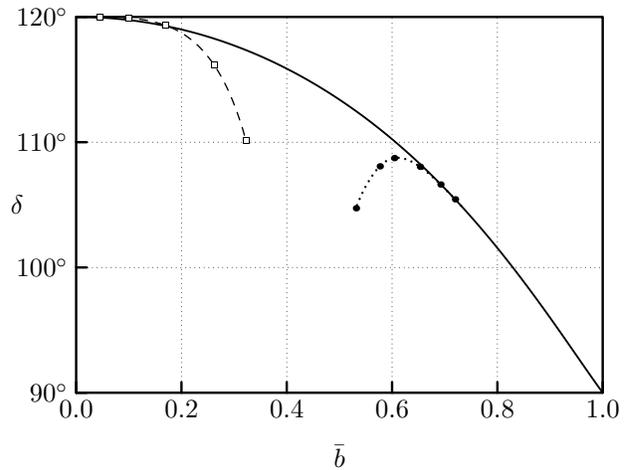}}
\caption{The angle $\delta$ of the cusp for the mass-shedding configurations in
 the Roche model (solid line) is compared with the numerical result for polytropic rings (dotted line)
 and polytropic spheroidal configurations (dashed line). The
 larger dots denote the ring configurations with $n=1/2$, $1$, $3/2$, $3$, $5$ and $n=7$, where smaller values of $n$
 correspond to smaller values of $\bar b$, and the white squares denote the spheroidal configurations
 with $n=1/2$, $1$, $2$, $3$ and $n=4$, where larger values of $n$ correspond to smaller values of $\bar b$.
 In the thin ring limit, where we expect that the Roche model provides arbitrarily good results to the full
 problem for `isothermal' rings ($n\to\infty$), we find $\delta\to 90^\circ$.
 \label{delta_bild}}
\end{figure}

Fig.~\ref{delta_bild} provides an interesting comparison between mass-shedding configurations for
polytropes and the Roche model. The angle $\delta$ of the cusp as given by \eqref{delta}
is plotted over the whole interval $\bar b\in[0,1]$ as a solid line. This curve is compared to
two distinct polytropic mass-shedding sequences generated by varying the polytropic index $n$.
The dotted line describes rings and merges together with that of the Roche model in the
thin ring limit, i.e.\ for $\bar b\to 1$. One the other hand, the dashed line describes (spheroidal)
stars and merges together with that of the Roche model for $\bar b\to 0$. The change in topologies
in the Roche model from toroidal to spheroidal takes place for $\bar b\approx 0.56$ as can be
calculated by solving for $\bar b$ in \eqref{V0o_V0i} with $A=0$ and $\alpha=1$.
\par

In summary, we can say that as with stars, the Roche model presented here for rings is
seen to yield a very good approximation to the full problem in many cases. Moreover, it presumably provides exact results
in the appropriate limit and offers new solutions that had not been found using perturbative approaches.

\section*{Acknowledgments}
   We are grateful to Reinhard Meinel for a careful reading of this paper.
   We also want to thank our referee, Yoshiharu Eriguchi, for his helpful comments.
   This research was funded in part by the Deutsche Forschungsgemeinschaft
   (SFB/TR7--B1).

%% file: Roche_appendix.tex
\section{The Standard Roche Model}\label{standard_Roche}

As mentioned in the introduction, the Roche model with a $1/r$-potential was
discussed by \citet{Roche73, ZN71, SS02} and provides a unique surface function
describing a mass-shedding star. To derive this function, we follow \citet{Meineletal08}
and begin by writing down equation \eqref{int_Euler_surf} with the potential $U_\text s$ of
a point mass
\begin{equation}\label{surface_equation}
  V_0 + \frac{GM}{\sqrt{\varrho^2+z_\text s(\varrho)^2}} + \frac{1}{2}\Omega^2\varrho^2 = 0.
\end{equation}
By considering the point $\varrho=0$, we can relate the constant $V_0$ to the
polar radius $z_\text p$
\begin{equation}
  V_0 =  -\frac{GM}{z_\text p}.
\end{equation}
For mass-shedding stars, the relation
\begin{equation}
  \left.\frac{\partial U}{\partial\varrho}\right|_{\varrho=\varrho_\text o,z=0} =  \varrho_\text o\Omega^2,
\end{equation}
where $\varrho_\text o$ is the equatorial radius, leads to
\begin{equation}\label{M/r^3_Omega^2}
  GM = \varrho_\text o^3\Omega^2,
\end{equation}
and the surface equation \eqref{surface_equation} becomes
\begin{equation}
  z_\text p\left(\frac{1}{\sqrt{\varrho^2+z^2}} + \frac{\varrho^2}{2 \varrho_\text o^3}\right) = 1.
\end{equation}
Evaluating this expression at the equator, $z=0$, tells us that for mass-shedding
fluids in the Roche model, the radius ratio
\begin{equation}\label{two-thirds}
\frac{z_\text p}{\varrho_\text o} = \frac{2}{3}
\end{equation}
follows. With this relationship, the curve describing the fluid's surface can be rewritten as
\begin{equation}\label{roche_surface_equation}
 z_\text s = \frac{\sqrt{4 \varrho_\text o^2-\varrho^2} \left(\varrho_\text o^2-\varrho^2\right)}{3\varrho_\text o^2-\varrho^2}.
\end{equation}
It then follows that
\begin{equation}\label{root_three}
 \lim_{z\downarrow0} \frac{\d z_\text s}{\d \varrho} = -\sqrt{3},
\end{equation}
which means the interior angle of the mass-shedding cusp is $120^\circ$ (see Fig.~\ref{delta_bild}
for $\bar b\to 0$).
\par

In general, the Roche model is expected to approach the full solution when the
matter is arbitrarily concentrated. For static polytropes, the full problem
leads to the Lane-Emden equation, which turns out to describe
arbitrarily concentrated matter if $n=5$. We shall now show the agreement between
these stars and the results given by the Roche model.
\par

In the static case, the Roche model gives
\begin{align}
 h = GM\left(\frac{1}{R} - \frac{1}{R_0}\right), \qquad R\in(0,R_0]
\end{align}
and thus
\begin{align}\label{dhdR}
 R^2\frac{\d h}{\d R} = -GM.
\end{align}
The solution to the full problem is
\begin{align}
  \mu_5 &=\mu_\text c\left(1+\frac{\xi^2}{3}\right)^{-5/2}, \qquad \xi\in[0,\infty) \intertext{and thus}
  h_5 &=6K\mu_5^{1/5},
\end{align}
where
\begin{align}
 \xi=\frac xl, \qquad l^2=\frac{3 K}{2\pi G\mu_\text c^{4/5}}
\end{align}
is the dimensionless radius and $x$ the true radius.
The surface extends out to infinity and since the regular portion of the $R$-interval, $R>0$, corresponds
to infinite values of the radius $x$, we consider
\begin{align}\label{dhdxi}
 \lim_{x\to\infty}x^2\frac{\d h_5}{\d x} = -6\sqrt{3}Kl\mu_\text c^{1/5}.
\end{align}
Calculating the mass for the full problem,
\begin{align}
 GM=4\pi G\int_0^\infty \mu_5x^2\,\d x = 4\pi\sqrt{3}l^3\mu_\text c=6\sqrt{3}Kl\mu_\text c^{1/5},
\end{align}
shows the equivalence of \eqref{dhdR} and \eqref{dhdxi}.

%% file: Roche.bbl
\begin{thebibliography}{}

\bibitem[\protect\citeauthoryear{Ansorg, Kleinw{\"a}chter \& Meinel}{Ansorg
  et~al.}{2003a}]{AKM03b}
Ansorg M.,  Kleinw{\"a}chter A.,    Meinel R.,  2003a, Astron.\ Astrophys.,
  405, 711

\bibitem[\protect\citeauthoryear{Ansorg, Kleinw{\"a}chter \& Meinel}{Ansorg
  et~al.}{2003b}]{AKM03}
Ansorg M.,  Kleinw{\"a}chter A.,    Meinel R.,  2003b, Mon.\ Not.\ R.\ Astron.\
  Soc., 339, 515

\bibitem[\protect\citeauthoryear{{Ansorg} \& {Petroff}}{{Ansorg} \&
  {Petroff}}{2005}]{AP05}
{Ansorg} M.,  {Petroff} D.,  2005, Phys.~Rev.~D, 72, 024019

\bibitem[\protect\citeauthoryear{{D}yson}{{D}yson}{1892}]{Dyson92}
{D}yson F.~W.,  1892, Philos.\ Trans.\ R.\ Soc.\ London, Ser. A, 184, 43

\bibitem[\protect\citeauthoryear{{D}yson}{{D}yson}{1893}]{Dyson93}
{D}yson F.~W.,  1893, Philos.\ Trans.\ R.\ Soc.\ London, Ser. A, 184, 1041

\bibitem[\protect\citeauthoryear{Eriguchi \& Hachisu}{Eriguchi \&
  Hachisu}{1985}]{EH85}
Eriguchi Y.,  Hachisu I.,  1985, Astron.\ Astrophys., 148, 289

\bibitem[\protect\citeauthoryear{Eriguchi \& Sugimoto}{Eriguchi \&
  Sugimoto}{1981}]{ES81}
Eriguchi Y.,  Sugimoto D.,  1981, Prog.\ Theor.\ Phys., 65, 1870

\bibitem[\protect\citeauthoryear{Fischer, Horatschek \& Ansorg}{Fischer
  et~al.}{2005}]{FHA05}
Fischer T.,  Horatschek S.,    Ansorg M.,  2005, Mon.\ Not.\ R.\ Astron.\ Soc.,
  364, 943

\bibitem[\protect\citeauthoryear{Hachisu}{Hachisu}{1986}]{Hachisu86}
Hachisu I.,  1986, Astrophys.\ J.\ Suppl.\ Ser., 61, 479

\bibitem[\protect\citeauthoryear{Kowalewsky}{Kowalewsky}{1885}]{Kowalewsky85}
Kowalewsky S.,  1885, Astronomische Nachrichten, 111, 37

\bibitem[\protect\citeauthoryear{Meinel, Ansorg, Kleinw{\"a}chter, Neugebauer
  \& Petroff}{Meinel et~al.}{2008}]{Meineletal08}
Meinel R.,  Ansorg M.,  Kleinw{\"a}chter A.,  Neugebauer G.,    Petroff D.,
  2008, Relativistic Figures of Equilibrium.
Cambridge University Press, Cambridge

\bibitem[\protect\citeauthoryear{Ostriker}{Ostriker}{1964a}]{Ostriker64}
Ostriker J.,  1964a, Astrophys.\ J., 140, 1056

\bibitem[\protect\citeauthoryear{Ostriker}{Ostriker}{1964b}]{Ostriker64b}
Ostriker J.,  1964b, Astrophys.\ J., 140, 1067

\bibitem[\protect\citeauthoryear{P{\"a}htz}{P{\"a}htz}{2007}]{Paehtz07}
P{\"a}htz T.,  2007, {U}n\-ter\-such\-un\-gen des {M}ass-{S}hedding-{L}imits
  ro\-tie\-ren\-der {F}l{\"u}s\-sig\-keit\-en in {N}ew\-ton\-scher und
  {E}in\-stein\-scher {G}ra\-vi\-ta\-ti\-ons\-theo\-rie, Di\-plom\-ar\-beit,
  Fried\-rich-Schil\-ler-Uni\-ver\-si\-t{\"a}t Jena

\bibitem[\protect\citeauthoryear{Petroff \& Horatschek}{Petroff \&
  Horatschek}{2008}]{PH08b}
Petroff D.,  Horatschek S.,  2008, Mon.\ Not.\ R.\ Astron.\ Soc., 389, 156

\bibitem[\protect\citeauthoryear{Poincar{\'e}}{Poincar{\'e}}{1885a}]{Poincare8%
5}
Poincar{\'e} H.,  1885a, Acta mathematica, 7, 259

\bibitem[\protect\citeauthoryear{Poincar{\'e}}{Poincar{\'e}}{1885b}]{Poincare8%
5b}
Poincar{\'e} H.,  1885b, C.\ R.\ Acad.\ Sci., 100, 346

\bibitem[\protect\citeauthoryear{Poincar{\'e}}{Poincar{\'e}}{1885c}]{Poincare8%
5c}
Poincar{\'e} H.,  1885c, Bull. Astr., 2, 109

\bibitem[\protect\citeauthoryear{Poincar{\'e}}{Poincar{\'e}}{1885d}]{Poincare8%
5d}
Poincar{\'e} H.,  1885d, Bull. Astr., 2, 405

\bibitem[\protect\citeauthoryear{Roche}{Roche}{1873}]{Roche73}
Roche {\'E}.,  1873, M\'em. de la section des sciences, Acad. des sciences et
  lettres de Montpellier, 1, 235

\bibitem[\protect\citeauthoryear{Shapiro \& Shibata}{Shapiro \&
  Shibata}{2002}]{SS02}
Shapiro S.~L.,  Shibata M.,  2002, Astrophys.\ J., 577, 904

\bibitem[\protect\citeauthoryear{Wong}{Wong}{1974}]{Wong74}
Wong C.~Y.,  1974, Astrophys.\ J., 190, 675

\bibitem[\protect\citeauthoryear{Zel'dovich \& Novikov}{Zel'dovich \&
  Novikov}{1971}]{ZN71}
Zel'dovich Y.~B.,  Novikov I.~D.,  1971, Relativistic Astrophysics.
Vol.~1, The University of Chicago Press, Chicago

\end{thebibliography}
